\documentclass[10pt,twocolumn,twoside] {IEEEtran}
\usepackage{times}
\usepackage{latexsym,amsmath,amstext,amsfonts,amssymb,epsfig}
\usepackage{cite}
\usepackage{latexsym}
\usepackage{color}

\title{Sparsity-Based STAP Design Based on Alternating Direction Method with Gain/Phase Errors}

\author{Zhaocheng~Yang,~\IEEEmembership{Member~IEEE}, Rodrigo C. de Lamare,~\IEEEmembership{Senior Member~IEEE}, and Weijian Liu~\IEEEmembership{Member~IEEE} 
\thanks{Z. Yang is with College of Information Engineering, Shenzhen University, Shenzhen, Guangdong, 518060, China. Email: yangzhaocheng@szu.edu.cn. R. C. de Lamare is with Department of Electronics, University of York, YO10 5DD, York, UK. Email: delamare@cetuc.puc-rio.br. W. Liu is with Wuhan Radar Academy, Wuhan 430019, China. Email: liuvjian@163.com}
\thanks{This work was supported in part by National Natural
Science Foundation of China under Grant 61401478,
the Science \& Technology Innovation
Project of Shenzhen under Grant JCYJ20160307112710376
and the Natural Science Foundation of
SZU under Grant 2016056.}
}
\begin{document}
%
\maketitle
\begin{abstract}
We present a novel sparsity-based space-time adaptive processing
(STAP) technique based on the alternating direction method to
overcome the severe performance degradation
caused by array gain/phase (GP) errors. The proposed algorithm
reformulates the STAP problem as a joint optimization
problem of the spatio-Doppler profile and
GP errors in both single and multiple snapshots, and
introduces a target detector using the reconstructed
spatio-Doppler profiles. Simulations
are conducted to illustrate the benefits of the proposed algorithm.
\end{abstract}
%
%
\section{Introduction}
\label{sec:intro}
Ground moving target detection (GMTD) in surveillance airborne radar is a crucial task for
many military and civilian applications. The technique of moving target indication (MTI) exploits
the differences of the Doppler frequencies between the targets and clutter
for the detection of targets \cite{JWard1994}. However, targets are often obscured by
the spreading Doppler spectrum of the clutter due to the moving
airborne platform, which leads to severe
detection performance degradation. Unlike MTI, space-time
adaptive processing (STAP) separates the target and  clutter from a
joint spatio-Doppler dimension, and exploits significantly more degrees of freedom (DoFs) than
MTI to mitigate clutter while preserving target energy\cite{JWard1994,Guerci2003}.

Because of the large number of space-time DoFs, full rank STAP
techniques have a slow convergence and requires a large number of
independent and identically distributed (IID) training snapshots
(e.g., twice the system DoFs according to the Reed-Mallett-Brennan
rule \cite{JWard1994,Guerci2003}), which is difficult to satisfy in
real scenarios, especially in nonhomogeneous environments
\cite{JWard1994,Guerci2003}. For example, to ensure a loss less than
$3$dB, it requires the ground with the range of $6$ kilometers to
satisfy the homogeneity for a range resolution of $30$ meters
(corresponding to a bandwidth of $5$MHz), $10$ antenna elements and
$10$ pulses. When the observation area is the city or the sea, the
above requirement is very difficult to be satisfied.
Reduced-dimension and reduced-rank methods
\cite{bar-ness,pados99,reed98,hua,goldstein,santos,qian,delamarespl07,xutsa,delamaretsp,kwak,xu&liu,delamareccm,wcccm,delamareelb,jidf,delamarecl,delamaresp,delamaretvt,jioel,delamarespl07,delamare_ccmmswf,jidf_echo,delamaretvt10,delamaretvt2011ST,delamare10,fa10,lei09,ccmavf,lei10,jio_ccm,ccmavf,stap_jio,zhaocheng,zhaocheng2,arh_eusipco,arh_taes,dfjio,rdrab,dcg_conf,dcg,dce,drr_conf,dta_conf1,dta_conf2,dta_ls,song,wljio,barc,jiomber,saalt}.,
including the principle-components (PC) methods
\cite{Haimovich1996}, joint-domain localized approach
\cite{Wang1994},  cross-spectral metric
 method \cite{JScott1997}, multistage Wiener filter \cite{JScott1999},
auxiliary-vector filtering \cite{Pados2001}, and joint
interpolation, decimation and filtering algorithm \cite{RuiSTAP2010},
have been developed to counteract the slow convergence of full rank STAP.
The parametric adaptive matched filter (PAMF) based on a multichannel
autoregressive model \cite{Roman2000} and sparse space-time
beamformers exploiting the sparsity of the received data and
filter weights \cite{ZcYangTSP2011, ZcYangIETSP2012} provide alternative solutions
to reduce the number of required IID snapshots. Recently, knowledge-aided
(KA) STAP techniques, which aim at exploiting environmental knowledge, have
been developed to enhance the detection performance especially
in the case of nonhomogeneous environments (see, e.g.,
\cite{R.Guerci2006,WangP2014,Jeong2015,ZYangIET2016,ZYangDSP2016,ALiu2016} and the references therein).
However, the exact form of prior knowledge is still
problem-dependent and hard to derive. Moreover, how to effectively
use the prior knowledge remains a topic for further investigation.
Direct data domain least-squares (D3-LS) STAP approaches use only the received
data in the cell under test (CUT) and require no training data,
thereby avoiding estimation distortion caused by different
statistics of the training data \cite{Sarkar2001, Mulgrew2010}. However, this benefit
comes at the cost of a reduced system DOFs
resulting in degraded performance.

More recently, motivated by compressive sensing techniques,
sparsity-based STAP has been applied to GMTD and its basic idea
is to formulate the observing scene with the target and
clutter \cite{SMARIA2006,IvanW2010,Jian2010,Sun2011, KeSun2011a,ZYangCSSP2016,Sen2012},
only the clutter \cite{ZYangIET2016,Sen2012,KeSun2009,ZcYangGRSL2012, ZcYangSP2013,ZcYangIEThomotopy, Sen2015,Qwu2016,ZWang2016,MaqEL2014}
or only the target \cite{Jason2010,Sen2012,Kim2014,Kim2015}
estimation problem as a sparse recovery/representation (SR) problem or
a low-rank matrix estimation problem \cite{Sen2015}.
Compared with conventional reduced-dimension and reduced-rank STAP algorithms, the sparsity-based
STAP algorithms provide high-resolution of the scene and exhibit
much better performance in a very small training
support, or even in a single snapshot. However, this approach
relies on the accuracy of the sparse model and suffers
performance degradation due to the model mismatches caused by array errors or the intrinsic clutter motion (ICM)\footnote{One
point worth mentioning is that standard STAP is relative robust because
these errors are captured in the adaptively estimated space-time covariance
matrix if assuming that they are constant over the coherent processing interval and
things are suitably narrowband. The only impact to detection is a potential loss of the output signal-to-interference-plus-noise ratio (SINR)
from steering vector mismatch.}.
A sparsity-based D3 STAP algorithm with the covariance
matrix taper (CMT) has been proposed to overcome the
model mismatches caused by the ICM \cite{ZYangCSSP2016}.
A sparsity-based STAP with
array gain/phase (GP) error self-calibration has been developed in \cite{MaqEL2014}, which
iteratively solves an SR problem and an LS
calibration problem. Since it requires to repeatedly
recover the scene in every iteration, the
computational complexity is high.

In this paper, we focus on the GMTD using the sparsity-based STAP
in the presence of array GP errors. We first
build the sparse measurement model by taking array GP errors into account.
Under the framework of the alternating direction method (ADM) \cite{JYang2011,Bilen2014},
we add a constraint to the array GP errors, and
transform the conventional sparsity-based STAP problem into
a joint optimization problem of the spatio-Doppler profile
and the array GP errors. Different from the conventional sparsity-based STAP,
the proposed algorithm simultaneously estimates the spatio-Doppler profile and
array GP errors resulting in adaptation to practical situations. Unlike the approach in \cite{MaqEL2014},
the proposed algorithm only requires the recovery procedure once, leading to
a reduced computational complexity.
Furthermore, we propose iterative approaches to solve the above
problem with both single snapshot and multiple snapshots.
A median constant false alarm (CFAR)
detector based on the reconstructed spatio-Doppler
profiles is developed for target detection. Finally, simulations are carried out to illustrate the performance and
computational complexity of the proposed
algorithm.

\begin{figure*}[htb]
\centering
\includegraphics[width=120mm]{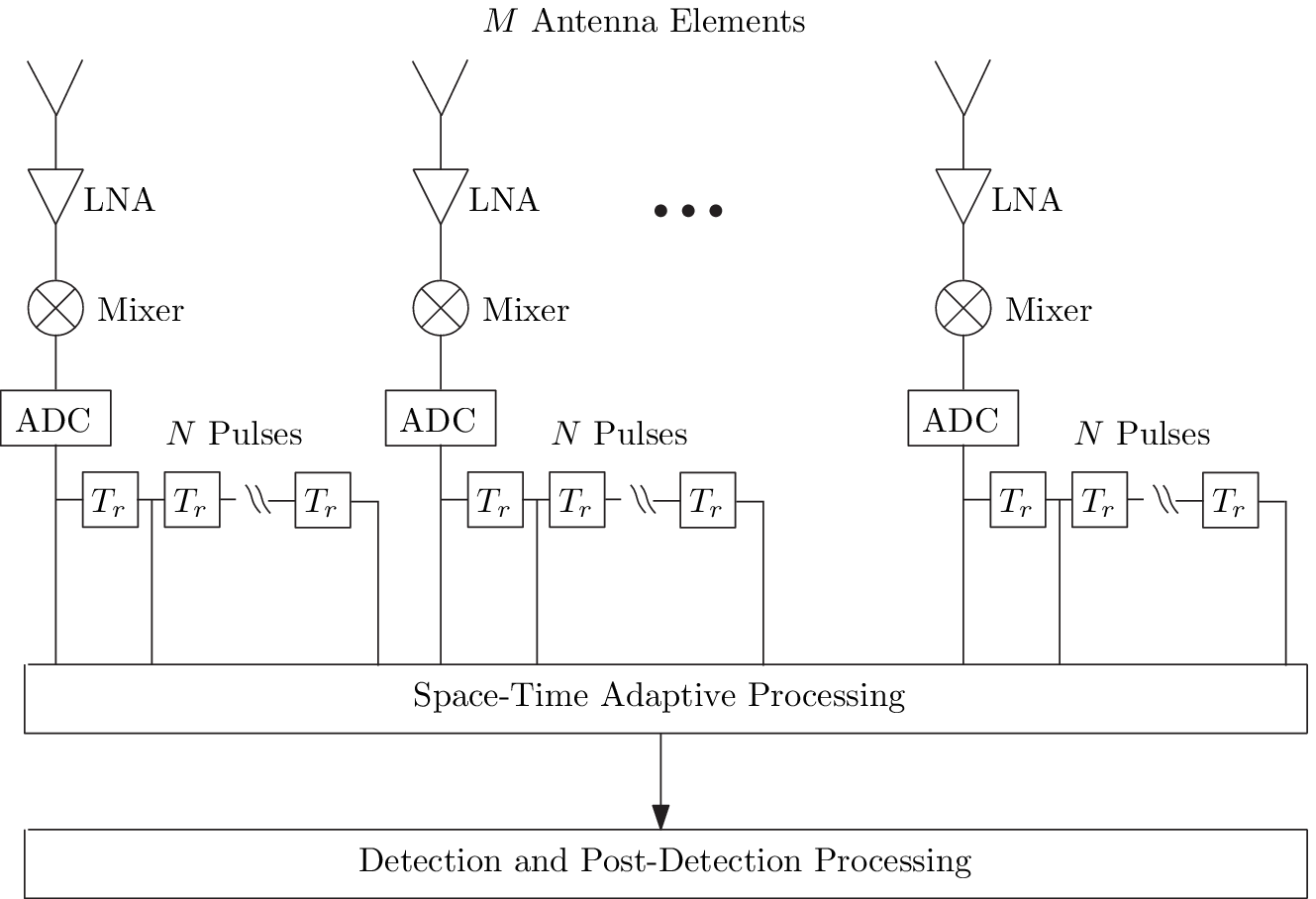}
\caption{\small
A general block diagram for a space-time processor.}\label{receiver}
\end{figure*}

The work is organized as follows. Section II introduces the STAP signal model
in the presence of array GP errors for airborne radar systems. Section III
builds the sparse signal model, and details the sparsity-based STAP using
the ADM framework. Simulated airborne radar data are used to evaluate
the performance of the proposed algorithm in Section IV. Section V
provides the summary and conclusions.

Notation: scalar quantities are denoted with italic typeface.
Lowercase boldface quantities denote vectors and uppercase
boldface quantities denote matrices. The operations of transposition,
complex conjugation, and conjugate transposition are denoted by
superscripts $T$, $\ast$, and $H$, respectively. The symbols
$\otimes$, $\odot$, $|\cdot|$, $\Re [\cdot]$, and $\|\cdot\|_p$ represent
the Kronecker product, Hadamard product, absolute value, real part
of the argument, and $l_p$-norm operation, respectively.

\section{Signal Model}
The airborne radar system under consideration employs a uniform linear
array (ULA) consisting of $M$ antenna elements with half wavelength inner
spacing $d_a=\lambda_c/2$ (where $\lambda_c$ is the carrier wavelength), as
shown in Fig.\ref{receiver}. Each element receives $N$
pulses in a coherent processing interval (CPI) with the pulse repetition interval (PRI) of $T_r$. At the receiver, each antenna element has
its own low-noise amplifier (LNA), mixer and AD converter (ADC). After these, all samples ($NM$) from the same CPI are combined for sequential processing. In general, the target detection problem for airborne radar
can be stated in the context of binary hypothesis
testing as given by
\begin{eqnarray}\label{eq1}
    H_0: {\bf x}& = &{\bf x}_u \nonumber\\
    H_1: {\bf x}& = &{\bf x}_t + {\bf x}_u,
\end{eqnarray}
where ${\bf x}_t$ is the target component, ${\bf x}_u$ is the
disturbance component, $H_0$ and $H_1$ denote target-absence and
target-presence, respectively. In reality, as the external
environment changes, such as temperature and humidity, the consistency of the amplifiers of the array multi-channel receivers is hard to keep. This inconsistency of the amplifiers is usually modeled as the array GP errors \cite{Guerci2003}. Hence, for a point target,
${\bf x}_t$ can be written as
\begin{eqnarray}\label{eq2}
    {\bf x}_t  = \alpha_t {\bf v}_d(f^t_d) \otimes \left({\bf c} \odot {\bf v}_s(f^t_s)\right)
     = \alpha_t{\bf C} {\bf v}(f^t_d, f^t_s)
\end{eqnarray}
where ${\bf c}=[c_1,\cdots,c_M]^T$ is the $M\times1$ array GP error vector,
${\bf C} = {\bf I}_{N} \otimes {\rm diag}({\bf c})$, ${\bf I}_N$ is an identity matrix
of size $N$,
${\bf v}_d(f^t_d)$ denotes the $N\times1$
temporal steering vector at the target Doppler frequency $f^t_d$,
${\bf v}_s(f^t_s)$ denotes the spatial steering vector in the direction
provided by the target spatial frequency $f^t_s$, and ${\bf v}(f^t_d, f^t_s)=
{\bf v}_d(f^t_d) \otimes {\bf v}_s(f^t_s)$ is the space-time steering vector
without array GP error. For the ULA, the steering vectors
${\bf v}_s(f^t_s)$ and ${\bf v}_d(f^t_d)$ are given by
\begin{eqnarray}\label{eq3}
    {\bf v}_s(f^t_s) = [1,  \exp\left(j2 \pi f^t_s\right),  \cdots,  \exp\left(j2(M-1)\pi f^t_s\right)]^T,
\end{eqnarray}
\begin{eqnarray}\label{eq4}
    {\bf v}_d(f^t_d) = [1,  \exp\left(j2 \pi f^t_d\right),  \cdots,  \exp\left(j2(N-1)\pi f^t_d\right)]^T.
\end{eqnarray}
The disturbance vector ${\bf x}_u$ is composed of the clutter
component ${\bf x}_c$ and the thermal noise component ${\bf n}$, i.e.,
${\bf x}_u = {\bf x}_c + {\bf n}$.
It is usually assumed that the clutter can be adequately approximated by
a summation of individual clutter patches over the iso-range of interest,
given by
\begin{eqnarray}\label{model1}
\begin{split}
    {\bf x}_{c} &= \sum^{N_c}_{k=1} \alpha_{c,k}
    {\bf v}_d(f^c_{d,k}) \otimes \left({\bf c} \odot {\bf v}_s(f^c_{s,k})\right) \\
    & = {\bf C} \sum^{N_c}_{k=1} \alpha_{c,k}
    {\bf v}(f^c_{d,k}, f^c_{s,k}),
\end{split}
\end{eqnarray}
where $N_c$ denotes the number of independent clutter patches,
$\alpha_{c,k}$ is the random complex amplitude of the $k$th clutter patch, and
$f^c_{s,k}$ and $f^c_{d,k}$ are the spatial and Doppler frequencies,
respectively, of the $k$th clutter patch. Moreover, it is assumed that the clutter amplitudes $\alpha_{c,k}$, $k=1,2,\cdots,N_c$, are IID mean-zero complex Gaussian random variables with variance $\sigma^2_{c,k}$. Hence, the corresponding clutter covariance matrix can be represented by
\begin{eqnarray}\label{model2}
    {\bf R}_{c} = {\bf C}\sum^{N_c}_{k=1} \sigma^2_{c,k}
    {\bf v}(f^c_{d,k}, f^c_{s,k}){\bf v}^H(f^c_{d,k}, f^c_{s,k}){\bf C}^H.
\end{eqnarray}
Additionally, we assume that the clutter spectral characteristics follow the local homogeneity. In the following, we will use this local homogeneity to estimate the clutter-plus-noise power level for target detection.
Since the thermal noise comes from the receiver electronics and is
added to the return after it passes through the antenna array, it is not affected
by the array errors. Here, we assume the thermal noise ${\bf n}$ is independent from
element to element and from pulse to pulse and follows the zero-mean Gaussian distribution
with covariance matrix ${\bf R}_n=\sigma^2_n{\bf I}_{NM}$.

\section{Proposed Sparsity-Based STAP in the Presence of Array GP Errors}
In this section, we first introduce the sparse signal model in the presence of array GP errors,
and then detail the sparsity-based STAP algorithm under the framework of ADM.

\subsection{Sparse Signal Model}
One notes that the clutter return in (\ref{model1}) is a function of the Doppler
frequency and spatial frequency. Let us discretize
the whole spatio-Doppler plane into a large number of
grid points (where $N_s=\rho_sM$, $N_d=\rho_d N$,
$\rho_s, \rho_d > 1$, $N_s$ and $N_d$ are the discretized number
of grid points along the spatial and Doppler frequencies, respectively) \cite{SMARIA2006}. A nonzero
element from any such grid point would suggest the presence of a scatterer
at that particular spatial and Doppler frequencies. We denote
the discretized spatial and Doppler frequencies of all grid points as
$\Psi = \left\{(f_{d,1}, f_{s,1}), (f_{d,1}, f_{s,2}), \cdots, (f_{d,N_d}, f_{s,N_s})\right\}$.
Therefore, the spatial and Doppler frequencies of the $N_c$ clutter
patches can be seen as a subset of $\Psi$, termed as $\Psi_c$.
Hence, similar to (\ref{model1}), the clutter return can be expressed by
\begin{eqnarray}\label{smodel1}
    {\bf x}_{c} = {\bf C}{\boldsymbol \Phi}{\boldsymbol \alpha}_c,
\end{eqnarray}
where ${\boldsymbol \alpha}_c=[\alpha_{1,1},\alpha_{1,2},\cdots, \alpha_{N_d,N_s}]^T$
denotes the $N_dN_s \times 1$ spatio-Doppler profile with nonzero elements
representing the clutter, and the $NM \times N_dN_s$ matrix ${\boldsymbol \Phi}$
is the over-complete space-time steering dictionary, as given by
\begin{eqnarray}\label{smodel2}
    {\boldsymbol \Phi}=[{\bf v}(f_{d,1}, f_{s,1}), {\bf v}(f_{d,1}, f_{s,2}),\cdots, {\bf v}(f_{d,N_d}, f_{s,N_s})],
\end{eqnarray}

The clutter sparsity can be understood from the
following two points: (a)
it is well known that, the relationship
between the Doppler frequency and the spatial frequency is a one-to-one mapping.
For example, the shapes of the clutter ridge are straight lines for the
side-looking ULA. Compared with the whole discretized place
(corresponding to the size of the set $\Psi$), the number of the
nonzero elements in the spatio-Doppler profile occupied
by the clutter (corresponding to the size of the
subset $\Psi_c$) is quite small.
(b) It is proved that for the
case of side-looking radar with a ULA, constant PRF, constant
platform velocity and no crab angle, there is a group of
space-time steering vectors (whose number is equivalent
to the clutter rank) that can approximately represent
the clutter subspace \cite{ZcYangGRSL2012}. That is to say
the clutter sparsity is much lower than the system DoFs and
far lower than $N_dN_s$ (since the clutter rank is much lower
than $NM$ and $NM \gg N_dN_s$). Similar conclusions are
also obtained by L. Bai \cite{BaiLin2013}. Moreover, according to
\cite{Wu2011}, the clutter rank can be estimated by counting
the number of resolution grids that are occupied by the
significant clutter spectrum components.
Therefore, there is a high degree of sparsity
of the clutter in the spatio-Doppler profile.

When a target is present in the CUT, corresponding
to $H_1$ hypothesis, the target's return is just like the response of
a nonzero element in the spatio-Doppler profile. If we
assume the target's spatial and Doppler frequencies are
from the grid points $\Psi$, then, the target return can be
written as
\begin{eqnarray}\label{smodel3.1}
    {\bf x}_t = {\bf C}{\boldsymbol \Phi}{\boldsymbol \alpha}_t,
\end{eqnarray}
where ${\boldsymbol \alpha}_t$ denotes the target amplitude.
Thus, the total return in the presence of target can be represented by
\begin{eqnarray}\label{smodel3}
    {\bf x} = {\bf x}_t + {\bf x}_c + {\bf n} = {\bf C}{\boldsymbol \Phi}{\boldsymbol \alpha} + {\bf n},
\end{eqnarray}
where ${\boldsymbol \alpha} = {\boldsymbol \alpha}_t + {\boldsymbol \alpha}_c$
represent the amplitudes from both the clutter and the target.
Because of the limitation of the number of
targets, it results in sparsity of the spatio-Doppler profile.

\subsection{Problem Formulation via the Framework of ADM}
For simplicity and convenience, we rewrite the expression
of (\ref{smodel3}) as
\begin{eqnarray}\label{smodel4}
    {\bf T}{\bf x} = {\boldsymbol \Phi}{\boldsymbol \alpha} + {\bf n}'.
\end{eqnarray}
where ${\bf T}={\bf I}_{N} \otimes {\rm diag}({\bf t})$, ${\bf t}=[t_1,\cdots,t_M]^T$
$t_m=c^{-1}_m$, $1\leq m \leq M$, and ${\bf n}'={\bf T}{\bf n}$. Here, we assume that
the unknown array GP errors are nonzeros.
Exploiting the sparsity of ${\boldsymbol \alpha}$,
the spatio-Doppler profile can be approximately
estimated by solving the so-called
basis pursuit denoising (BPDN) problem, described by
\begin{eqnarray}\label{admm1}
    \min_{{\boldsymbol \alpha}} \left\|{\boldsymbol \alpha}\right\|_1
    + \frac{1}{2\rho} \left\|{\bf T}{\bf x} - {\boldsymbol \Phi}{\boldsymbol \alpha}\right\|^2_2,
\end{eqnarray}
where $\rho>0$ is the positive regularization parameter that provides a trade-off between the
sparsity and total squared error. With an auxiliary variable
${\bf r} ={\bf T}{\bf x} - {\boldsymbol \Phi}{\boldsymbol \alpha}$, the above BPDN problem can be reformulated as
\begin{eqnarray}\label{admm2}
\begin{split}
    & \min_{{\boldsymbol \alpha},{\bf r}} \left\|{\boldsymbol \alpha}\right\|_1
    + \frac{1}{2\rho} \left\|{\bf r}\right\|^2_2 \\
    &  {\rm s.t.} \quad
{\boldsymbol \Phi}{\boldsymbol \alpha} + {\bf r} = {\bf T}{\bf x}
\end{split}
\end{eqnarray}

Then, the augmented Lagrangian function of this problem is given by \cite{JYang2011,Bilen2014}
\begin{eqnarray}\label{admm3}
\begin{split}
    & \min_{{\boldsymbol \alpha},{\bf r},{\boldsymbol \lambda},{\bf t}} \mathcal{L}'\left({\boldsymbol \alpha},{\bf r},{\boldsymbol \lambda},{\bf t}\right)=
    \min_{{\boldsymbol \alpha},{\bf r},{\boldsymbol \lambda},{\bf t}} \left\|{\boldsymbol \alpha}\right\|_1
    + \frac{1}{2\rho} \left\|{\bf r}\right\|^2_2 \\
    &  \quad -
    \Re\left\{{\boldsymbol \lambda}^H\left({\boldsymbol \Phi}{\boldsymbol \alpha} + {\bf r} - {\bf T}{\bf x}\right)\right\} + \frac{\beta}{2}\left\|{\boldsymbol \Phi}{\boldsymbol \alpha} + {\bf r} - {\bf T}{\bf x}\right\|^2_2
\end{split}
\end{eqnarray}
where ${\boldsymbol \lambda} \in \mathbb{C}^{NM}$ is a Lagrange multiplier and
$\beta>0$ is a penalty parameter. Note that the matrix ${\bf T}$ of (\ref{admm3})
depends on the array GP error vector ${\bf c}$ (${\bf t}$), which is unknown
and should be estimated from the data.
Given the snapshot ${\bf x}$ and the over-complete
space-time steering dictionary ${\boldsymbol \Phi}$,
we can obtain the spatio-Doppler profile ${\boldsymbol \alpha}$,
the auxiliary variable ${\bf r}$ and the array
GP error vector ${\bf c}$ by applying alternating minimization to solve
(\ref{admm3}).

With the above formulation, we observe that the problem (\ref{admm3})
is an unconstrained convex optimization problem. However, it is trivially satisfied for zeros of vectors ${\bf t}$ and ${\boldsymbol \alpha}$. To avoid
this trivial solution, we introduce a convex normalization constraint
$\sum^M_{m=1}t_m=\varsigma$, where $\varsigma \in \mathbb{C}$ is an
arbitrary constant scale. Therefore, the cost function $\mathcal{L}'\left({\boldsymbol \alpha},{\bf r},{\boldsymbol \lambda},{\bf t}\right)$ in problem (\ref{admm3})
can be rewritten as
\begin{eqnarray}\label{admm4}
\begin{split}
    & \mathcal{L}\left({\boldsymbol \alpha},{\bf r},{\boldsymbol \lambda},{\bf t}\right) = \left\|{\boldsymbol \alpha}\right\|_1
    + \frac{1}{2\rho} \left\|{\bf r}\right\|^2_2 - \Re\left\{\gamma^\ast\left(\sum^M_{m=1}t_m-\varsigma\right)\right\}\\
    &  \quad -
    \Re\left\{{\boldsymbol \lambda}^H\left({\boldsymbol \Phi}{\boldsymbol \alpha} + {\bf r} - {\bf T}{\bf x}\right)\right\} + \frac{\beta}{2}\left\|{\boldsymbol \Phi}{\boldsymbol \alpha} + {\bf r} - {\bf T}{\bf x}\right\|^2_2
\end{split}
\end{eqnarray}
where $\gamma$ is a Lagrange multiplier. The actual array GP error vector is recovered
after the optimization using $c_m = 1/t_m$, $m=1,\cdots,M$. One should also
note that the estimated array GP error vector scales to the true one because
of the constant scale $\varsigma$ in the constraint.

\subsection{Jointly Iterative Estimation of the Spatio-Doppler Profile and Array GP Error}
In this subsection, we estimate the spatio-Doppler profile and array
GP error vector iteratively. For ${\boldsymbol \alpha}={\boldsymbol \alpha}^p$,
${\boldsymbol \lambda}={\boldsymbol \lambda}^p$, and ${\bf t}={\bf t}^p$ fixed ($()^p$ denotes the
$p$th iteration), the minimizer of (\ref{admm4})
with respect to ${\bf r}^\ast$ is given by
\begin{eqnarray}\label{admm5}
    {\bf r}^{p+1} = \frac{\rho\beta}{1+\rho\beta} \left(\frac{{\boldsymbol \lambda}^p}{\beta}-
    {\boldsymbol \Phi}{\boldsymbol \alpha}^p + {\bf T}^p{\bf x}\right).
\end{eqnarray}
Similarly, for ${\bf r}={\bf r}^{p+1}$, ${\boldsymbol \lambda}={\boldsymbol \lambda}^p$,
and ${\bf t}={\bf t}^p$ fixed, the minimization of (\ref{admm4}) with respect to ${\boldsymbol \alpha}^\ast$
is equivalent to
\begin{eqnarray}\label{admm6}
    \min_{{\boldsymbol \alpha}} \left\|{\boldsymbol \alpha}\right\|_1
    + \frac{\beta}{2}\left\|{\boldsymbol \Phi}{\boldsymbol \alpha} + {\bf r}^{p+1} - {\bf T}^p{\bf x} - \frac{{\boldsymbol \lambda}^p}{\beta}\right\|^2_2.
\end{eqnarray}
Then, the solution of the problem (\ref{admm6}) can be approximately given by \cite{JYang2011,Bilen2014}
\begin{eqnarray}\label{admm7}
\begin{split}
    {\boldsymbol \alpha}^{p+1} & = {\rm soft}\left({\boldsymbol \alpha}^p - \tau {\bf g}^p, \frac{\tau}{\beta}\right)\\
    & = \max\left\{\left|{\boldsymbol \alpha}^p - \tau {\bf g}^p\right|-\frac{\tau}{\beta}, 0\right\}\frac{{\boldsymbol \alpha}^p - \tau {\bf g}^p}{\left|{\boldsymbol \alpha}^p - \tau {\bf g}^p\right|},
\end{split}
\end{eqnarray}
where all the operations in (\ref{admm7}) are performed component-wise (usually known as
shrinkage), $\frac{0}{0}=0$, $\tau>0$ is a proximal parameter and
\begin{eqnarray}\label{admm8}
    {\bf g}^p = {\boldsymbol \Phi}^H\left({\boldsymbol \Phi}{\boldsymbol \alpha}^p + {\bf r}^{p+1} - {\bf T}^p{\bf x} - \frac{{\boldsymbol \lambda}^p}{\beta}\right).
\end{eqnarray}

Given ${\bf r}={\bf r}^{p+1}$, ${\boldsymbol \lambda}={\boldsymbol \lambda}^p$
and ${\boldsymbol \alpha}={\boldsymbol \alpha}^{p+1}$,
the minimization of (\ref{admm4}) with respect to
${\bf t}^\ast$ can be simplified as
\begin{eqnarray}\label{admm9}
\begin{split}
    \min_{{\bf t}} & \frac{\beta}{2}\left\|{\boldsymbol \Phi}{\boldsymbol \alpha}^{p+1} + {\bf r}^{p+1} - {\bf T}{\bf x} - \frac{{\boldsymbol \lambda}^p}{\beta}\right\|^2_2 \\
    & \quad- \Re\left\{\gamma^\ast\left(\sum^M_{m=1}t_m-\varsigma\right)\right\}
\end{split}
\end{eqnarray}
For simplicity, we denote ${\bf z}^p = {\boldsymbol \Phi}{\boldsymbol \alpha}^{p+1} + {\bf r}^{p+1} - \frac{{\boldsymbol \lambda}^p}{\beta}$, ${\bf T}{\bf x}={\bf Q}{\bf t}$ and ${\bf Q} = {\rm diag}({\bf x})({\bf 1}_N \otimes {\bf I}_M)$.
Then, (\ref{admm9}) can be rewritten with the form of
\begin{eqnarray}\label{admm10}
    \min_{{\bf t}} & \frac{\beta}{2}\left\|{\bf z}^p - {\bf Q}{\bf t}\right\|^2_2 - \Re\left\{\gamma^\ast\left(\sum^M_{m=1}t_m-\varsigma\right)\right\}.
\end{eqnarray}

By taking the gradient of the cost function in problem (\ref{admm10}) with respect to
${\bf t}^\ast$ and $\gamma^\ast$, equating the terms to zero, and solving for ${\bf t}$, we obtain
(the detailed derivations are given in Appendix \ref{secapp1})
\begin{eqnarray}\label{admm11}
    {\bf t}^{p+1} = \left[\frac{b_1 + \gamma}{a_1}, \frac{b_2 + \gamma}{a_2}, \cdots, \frac{b_M + \gamma}{a_M}\right]^T,
\end{eqnarray}
where $b_m$, $a_m$ and $\gamma$, ($m=1,2,\cdots,M$) are defined by (\ref{app4}),
(\ref{app5}) and (\ref{app6}) in Appendix \ref{secapp1}.

Finally, minimizing (\ref{admm4}) with respect to ${\boldsymbol \lambda}^\ast$, we obtain
the update of the multiplier ${\boldsymbol \lambda}$ as
\begin{eqnarray}\label{admm12}
    {\boldsymbol \lambda}^{p+1} = {\boldsymbol \lambda}^p - \beta\left({\boldsymbol \Phi}{\boldsymbol \alpha}^{p+1} + {\bf r}^{p+1} - {\bf T}^{p+1}{\bf x}\right).
\end{eqnarray}

In short, the proposed approach iteratively updates
(\ref{admm5}), (\ref{admm7}), (\ref{admm11}) and (\ref{admm12})
to obtain estimates of the spatio-Doppler profile and
array GP error vector.

\subsection{Application to Multiple Snapshots}
It is reasonable to suppose that the array GP errors are identical for
different snapshots from adjacent range bins in the same CPI. By using multiple snapshots, we can expect to improve the accuracy of the estimated array GP errors and spatio-Doppler profiles. In the following, we apply the
proposed ADM algorithm to the multiple snapshots case. For $L$ snapshots,
${\bf x}_l$, $l=1,2,\cdots,L$, the problem of (\ref{admm4}) can be reformulated
as
\begin{eqnarray}\label{admmm1}
\begin{split}
    & \min_{{\boldsymbol \alpha}_l,{\bf r}_l,{\bf t}_l} \sum^L_{l=1}\left\|{\boldsymbol \alpha}_l\right\|_1
    + \frac{1}{2\rho} \sum^L_{l=1}\left\|{\bf r}_l\right\|^2_2 \\
    &  \quad -
    \Re\left\{\sum^L_{l=1}{\boldsymbol \lambda}^H_l\left({\boldsymbol \Phi}{\boldsymbol \alpha}_l + {\bf r}_l - {\bf T}{\bf x}_l\right)\right\} \\
     &  + \frac{\beta}{2}\sum^L_{l=1}\left\|{\boldsymbol \Phi}{\boldsymbol \alpha}_l + {\bf r}_l - {\bf T}{\bf x}_l\right\|^2_2 - \Re\left\{\gamma^\ast\left(\sum^M_{m=1}t_m-\varsigma\right)\right\},
\end{split}
\end{eqnarray}
where ${\boldsymbol \alpha}_l$, ${\bf r}_l$ and ${\boldsymbol \lambda}_l$ denote
the corresponding variables of the $l$th snapshot.

Let us define ${\boldsymbol \Upsilon} = [{\boldsymbol \alpha}_1,{\boldsymbol \alpha}_2,\cdots,{\boldsymbol \alpha}_L]$,
${\boldsymbol \Gamma} = [{\bf r}_1,{\bf r}_2,\cdots,{\bf r}_L]$,
${\boldsymbol \Lambda} = [{\boldsymbol \lambda}_1,{\boldsymbol \lambda}_2,\cdots,{\boldsymbol \lambda}_L]$, and
${\bf X} = [{\bf x}_1, {\bf x}_2, \cdots, {\bf x}_L]$.
Similar to the derivations in the previous subsection, we can subsequently obtain
the updates of ${\boldsymbol \Upsilon}$, ${\boldsymbol \Gamma}$ and ${\boldsymbol \Lambda}$
as
\begin{eqnarray}\label{admmm6}
    {\boldsymbol \Gamma}^{p+1} = \frac{\rho\beta}{1+\rho\beta} \left(\frac{{\boldsymbol \Lambda}^p}{\beta}-
    {\boldsymbol \Phi}{\boldsymbol \Upsilon}^p + {\bf T}^p{\bf X}\right),
\end{eqnarray}
\begin{eqnarray}\label{admmm7}
    {\boldsymbol \Upsilon}^{p+1} = {\rm soft}\left({\boldsymbol \Upsilon}^p - \tau {\bf G}^p, \frac{\tau}{\beta}\right),
\end{eqnarray}
and
\begin{eqnarray}\label{admmm8}
    {\boldsymbol \Lambda}^{p+1} = {\boldsymbol \Lambda}^p - \beta\left({\boldsymbol \Phi}{\boldsymbol \Upsilon}^{p+1} + {\boldsymbol \Gamma}^{p+1} - {\bf T}^{p+1}{\bf X}\right),
\end{eqnarray}
where
\begin{eqnarray}\label{admm9}
    {\bf G}^p = {\boldsymbol \Phi}^H\left({\boldsymbol \Phi}{\boldsymbol \Upsilon}^p + {\boldsymbol \Gamma}^{p+1} - {\bf T}^p{\bf X} - \frac{{\boldsymbol \Upsilon}^p}{\beta}\right).
\end{eqnarray}

The update of the vector ${\bf t}$
with multiple snapshots case can be represented by
\begin{eqnarray}\label{admmm2}
    {\bf t}^{p+1} = \left[\frac{\sum^L_{l=1}b_{l,1} + \tilde{\gamma}}{\sum^L_{l=1}a_{l,1}},  \cdots, \frac{\sum^L_{l=1}b_{l,M} + \tilde{\gamma}}{\sum^L_{l=1}a_{l,M}}\right]^T,
\end{eqnarray}
where
\begin{eqnarray}\label{admmm3}
     b_{l,m} = \sum^L_{l=1}\sum^N_{n=1} x^\ast_{l,(n-1)M+m}z^p_{l,(n-1)M+m},
\end{eqnarray}
\begin{eqnarray}\label{admmm4}
     a_{l,m} = \sum^L_{l=1}\sum^N_{n=1} \left|x_{l,(n-1)M+1}\right|^2,
\end{eqnarray}
\begin{eqnarray}\label{admmm5}
     \tilde{\gamma} = \frac{\varsigma - \sum^M_{m=1}\frac{\sum^L_{l=1}b_{l,m}}{\sum^L_{l=1}a_{l,m}}}{\sum^M_{m=1}\frac{1}{\sum^L_{l=1}a_m}},
\end{eqnarray}
and
\begin{eqnarray}\label{admmm10}
\begin{split}
     {\bf Z}^p = [{\bf z}^p_1,{\bf z}^p_2,\cdots,{\bf z}^p_L] ={\boldsymbol \Phi}{\boldsymbol \Upsilon}^{p+1} + {\boldsymbol \Gamma}^{p+1} - \frac{{\boldsymbol \Lambda}^p}{\beta}.
\end{split}
\end{eqnarray}
Moreover, we detail the proposed ADM algorithm for
jointly iterative estimation of the spatio-Doppler profile and
the array GP error vector (shortened as JIE-ADM) in Table \ref{JIE-ADM}.

\begin{table}[!ht]
  \centering
  \caption{The Proposed JIE-ADM Algorithm}\label{JIE-ADM}
  \begin{tabular}{l}
  \hline
  \textbf{Initialization:}\\
  ${\boldsymbol \alpha}^0_l = {\bf 0}_{N_dN_s}$, ${\boldsymbol \lambda}^0_l = {\bf 0}_{NM}$,
  $l=1,\cdots,L$, \\
  ${\bf t}^0 = {\bf 1}_M$, ${\bf T}^0={\bf I}_{N}\otimes {\rm diag}({\bf t}^0)$, $p=0$\\
  \hline
  \textbf{Repeat}\\
  1\quad ${\boldsymbol \Gamma}^{p+1} = \frac{\rho\beta}{1+\rho\beta} \left(\frac{{\boldsymbol \Lambda}^p}{\beta}-
    {\boldsymbol \Phi}{\boldsymbol \Upsilon}^p + {\bf T}^p{\bf X}\right)$,\\
  2\quad ${\bf G}^p = {\boldsymbol \Phi}^H\left({\boldsymbol \Phi}{\boldsymbol \Upsilon}^p + {\boldsymbol \Gamma}^{p+1} - {\bf T}^p{\bf X} - \frac{{\boldsymbol \Upsilon}^p}{\beta}\right)$,\\
  3\quad ${\boldsymbol \Upsilon}^{p+1} = {\rm soft}\left({\boldsymbol \Upsilon}^p - \tau {\bf G}^p, \frac{\tau}{\beta}\right)$,\\
  4\quad Update ${\bf Z}^p$, $b_{l,m}$ $a_{l,m}$ and $\tilde{\gamma}$ by (\ref{admmm10}), (\ref{admmm3}) (\ref{admmm4}), and (\ref{admmm5}),\\
  5\quad ${\bf t}^{p+1} = \left[\frac{\sum^L_{l=1}b_{l,1} + \tilde{\gamma}}{\sum^L_{l=1}a_{l,1}},  \cdots, \frac{\sum^L_{l=1}b_{l,M} + \tilde{\gamma}}{\sum^L_{l=1}a_{l,M}}\right]^T$\\
  6\quad ${\boldsymbol \Lambda}^{p+1} = {\boldsymbol \Lambda}^p - \beta\left({\boldsymbol \Phi}{\boldsymbol \Upsilon}^{p+1} + {\boldsymbol \Gamma}^{p+1} - {\bf T}^{p+1}{\bf X}\right)$, \\
  \textbf{Until} $\frac{\sum^L_{l=1}\|{\boldsymbol \alpha}^p_l-{\boldsymbol \alpha}^{p+1}_l\|_2}{\sum^L_{l=1}\|{\boldsymbol \alpha}^{p+1}_l\|_2} \leq \zeta$\\
  \hline
   \end{tabular}
\end{table}

\begin{figure*}[t]
\centering
\includegraphics[width=150mm]{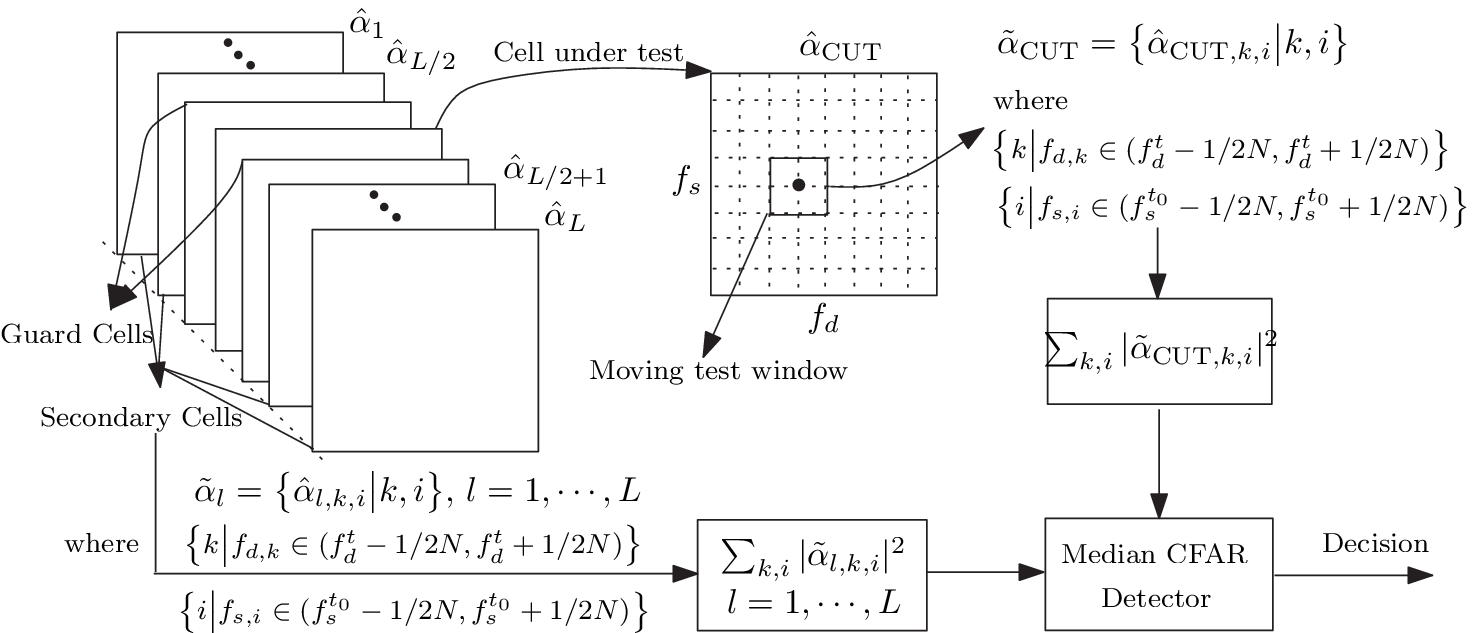}
\caption{\small
The procedure of the median CFAR detector.}\label{MCFARDetector}
\end{figure*}

\subsection{Target Detection}
Since the main purpose of this paper is
to improve the spatio-Doppler profiles' estimation by using the
joint estimation approach, in this subsection, we only consider using
a simple detector for illustration purposes. Furthermore, following the ideas in \cite{Jian2010}, we focus on target detection based on
the estimated spatio-Doppler profiles.
It is reasonable to assume that the reconstructed clutter
peaks at a few (say, $10$) adjacent range bins are
nearly the same, and the target peaks are not so ``dense"
in range \cite{Jian2010}. Therefore, as shown in Fig.\ref{MCFARDetector},
we first exclude some snapshots around the CUT (namely the guard cells)
for avoiding target canceling, and then perform a
moving test window to the estimated spatio-Doppler
profile $\hat{\boldsymbol \alpha}_{\rm CUT}$ at the CUT
with the size of a spatio-Doppler
resolution cell (i.e., $1/N$ and $1/M$
for the Doppler and spatial frequencies, respectively).
When we conduct the detection procedure,
we should determine the presence/absence of the target
for every single angle and every single Doppler frequency.
Since the airborne radar transmitter usually keeps a high
gain in the observing angle for a CPI, we only require to
conduct the detection procedure by fixing the spatial frequency
of detection to the main-lobe $f^{t_0}_{s}$ and varying
the Doppler frequency within a set of possible values.
Specifically, for a possible target Doppler frequency $f^t_d$, the range of
the moving test window is $(f^t_d-1/2N, f^t_d+1/2N)$ and
$(f^{t_0}_{s}-1/2M, f^{t_0}_{s}+1/2M)$.
Then, we pick out the elements that belong to the moving test window
from $\hat{\boldsymbol \alpha}_{\rm CUT}$, and arrange them into
a new vector $\tilde{\boldsymbol \alpha}_{\rm CUT}$, as given by
\begin{eqnarray}\label{det1}
\begin{split}
    \tilde{\boldsymbol \alpha}_{\rm CUT} &= \bigg\{\hat{\alpha}_{{\rm CUT},k,i} \bigg|f_{d,k} \in \left(f^t_{d}-\frac{1}{2N}, f^t_{d}+\frac{1}{2N}\right), \\
    & \qquad \qquad f_{s,i} \in \left(f^{t_0}_s-\frac{1}{2M}, f^{t_0}_s+\frac{1}{2M}\right)\bigg\},
\end{split}
\end{eqnarray}
Similarly, for the same moving test window, we form $L$ secondary samples
$\tilde{\boldsymbol \alpha}_{l}$, $l=1,2,\cdots,L$, by
\begin{eqnarray}\label{det2}
\begin{split}
    \tilde{\boldsymbol \alpha}_l &= \bigg\{\hat{\alpha}_{l,k,i} \bigg|f_{d,k} \in \left(f^t_{d}-\frac{1}{2N}, f^t_{d}+\frac{1}{2N}\right), \\
    & \qquad \qquad f_{s,i} \in \left(f^{t_0}_s-\frac{1}{2M}, f^{t_0}_s+\frac{1}{2M}\right)\bigg\},
\end{split}
\end{eqnarray}

Due to the estimation errors or the
discretized errors, the target energy might be not just
concentrated at a single discretized spatio-Doppler grid point.
Therefore, we select the sum
value of absolute elements in $\tilde{\boldsymbol \alpha}_{\rm CUT}$ as the test statistic
for each spatio-Doppler resolution cell.
Similar operations are carried out for the secondary samples
$\tilde{\boldsymbol \alpha}_{l}$, $l=1,\cdots,L$, which are
used to generate the background clutter-plus-noise level. Finally,
we use a median CFAR detector with the form of \cite{Bergin2004}
\begin{eqnarray}\label{det3}
    20\log \vartheta_{\rm CUT} - 20 \log {\rm median}\left(\vartheta_{l}\right) {{H_1\atop>}\atop {<\atop{H_0}}}\xi,
\end{eqnarray}
where $l=1,2,\cdots,L$, $\xi$ is the threshold scalar, ${\rm median}(\cdot)$
yields the median value of samples in the parentheses, $\log$
represents the logarithm taking $10$ as the base,
and $\vartheta_{\rm CUT}$ and $\vartheta_{l}$ are given by
\begin{eqnarray}\label{det4}
\begin{split}
    \vartheta_{\rm CUT} &= \sum_{k,i} \left|\tilde{\alpha}_{{\rm CUT},k,i}\right|, \\
    \vartheta_{l} &= \sum_{k,i} \left|\tilde{\alpha}_{l,k,i}\right|.
\end{split}
\end{eqnarray}

\section{Performance Assessment}
In this section, we evaluate the performance of the proposed
JIE-ADM algorithm in terms of
qualities of the reconstructed spatio-Doppler profiles
and the probability of detection (PD) using simulated
data. For comparison purposes, we also
show the performance of the proposed JIE-ADM algorithm, conventional D3-LS STAP \cite{Sarkar2001},
ADM \cite{JYang2011}, basis pursuit using interior-point method (BP-IPM) \cite{cvx}, and IAA \cite{Jian2010}
and ADMT (using the ADM reconstructs the spatio-Doppler
profile with the known array GP errors).

\label{simu}
The parameters of
the simulated radar platform are shown in
Table \ref{paraRadar}. In addition, for each range bin, the $[-\pi/2, \pi/2]$
AOA interval was divided into $361$ clutter patches, whose single channel, single
pulse clutter-to-noise ratio (CNR) is $30$dB. The thermal noise power for each
channel and each pulse is set to unit. The gain error and phase error are
both assumed to follow a uniform distribution \cite{MaqEL2014,ALiu2016}. Specifically, we can denote the $m$th
entry of the array GP error vector as $c_m=(1+\epsilon_m) e^{j\phi_m}$, $m=1,2,\cdots,M$, where
$\epsilon_m$ and $\phi_m$ follow a uniform distribution within
$[-\epsilon_{\rm max}, \epsilon_{\rm max}]$ and $[-\phi_{\rm max}, \phi_{\rm max}]$,
respectively.

Additionally, in the following simulations,
for the JIE-ADM and ADM algorithms, $\beta=0.1$, $\rho=0.01$,
$\zeta=10^{-4}$ and the maximum iteration number $500$. For the BP-IPM, the noise allowance
parameter is set to $10^{-3}$ and the maximum iteration number $500$.
For the IAA, the stopping criterion is decided by
both the preset limit relative change of the
solutions between two adjacent iterations $10^{-4}$ and the maximum iteration number $20$.
Moreover, the whole spatio-Doppler plane is discretized into $ N_d \times N_s=5N \times 5M $ grid points
for all algorithms.

\begin{table}[!ht]
  \centering
  \caption{Parameters of Airborne Radar System}\label{paraRadar}
  \begin{tabular}{ll}
  \hline
  Parameter & Value \\
   \hline
   Antenna array & Side-looking ULA\\
   Antenna array spacing & $\lambda_c/2$\\
   Carrier frequency & $1.24$GHz\\
   Transmit taper & Uniform\\
   PRF & $1984$Hz\\
   Platform velocity & $100$m/s\\
   Platform height & $3000$ m\\
   Antenna elements number & $10$\\
   Pulse number in one CPI & $10$\\
   \hline
   \end{tabular}
\end{table}


\begin{figure}[t]
\centering
  \includegraphics[width=53mm]{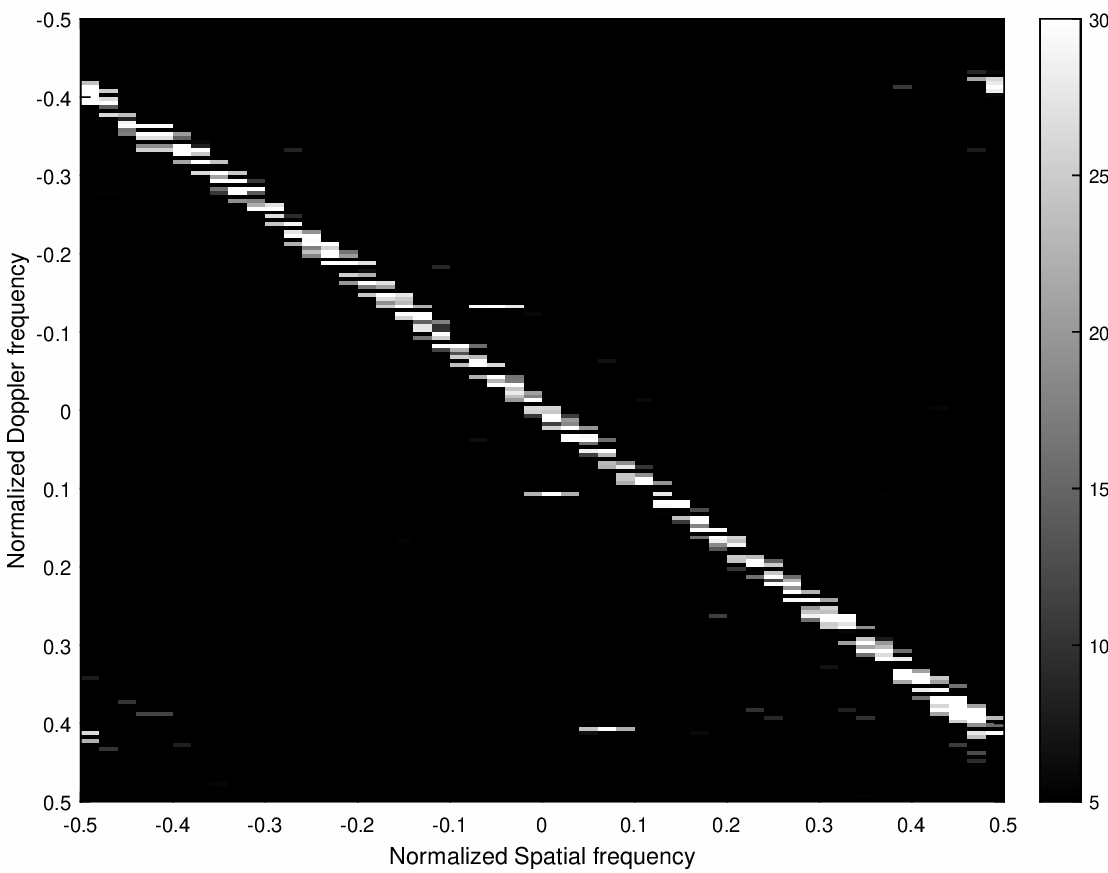}
  \caption{The reconstructed spatio-Doppler profiles with array GP
  errors $\epsilon_{\rm max}=0.1$ and $\phi_{\rm max}=0.1 \pi$ when the number of transmitted pulses is $100$.}
  \label{sdprofile2}
\end{figure}

In the first example, we focus on the spatio-Doppler profile reconstructions
considering different cases of array GP errors: case 1, no array GP error,
i.e., $\epsilon_{\rm max}=0$ and $\phi_{\rm max}=0$; case 2, $\epsilon_{\rm max}=0.05$ and $\phi_{\rm max}=0.05 \pi$;
case 3, $\epsilon_{\rm max}=0.1$ and $\phi_{\rm max}=0.1 \pi$.
In addition, we assume that there are three targets in the boresight
at the range bin of interest: target 1 with the normalized Doppler frequency
equal to $-0.13$ and the input signal-to-noise ratio (SNR) set to $0.2$dB; target 2 with the normalized Doppler frequency
equal to $0.11$ and the target's input SNR set to $-3.8$dB; and target 3 with the normalized Doppler frequency equal to
$0.41$ and the target's input SNR set to $-3.8$dB. Here, we set a larger target's input SNR for target 1
because it stands for a slow target and is not well recovered when in
small input SNR.
As shown in Fig.\ref{sdprofile}, we see that the spatio-Doppler profiles
can be well reconstructed for the ADMT when the array GP errors are known. It
is also observed that more and more pseudo peaks are present in the spatio-Doppler
profiles using the ADM, BP-IPM and IAA algorithms, as the increase of the array GP errors.
On the contrary, the spatio-Doppler profiles using the proposed JIE-ADM
algorithm keep nearly the same qualities as those using the ADMT.

To better illustrate the performance of the proposed algorithm,
we conduct simulations with a large number of pulses (i.e., $100$). The reconstructed spatio-Doppler profiles with array GP errors $\epsilon_{\rm max}=0.1$ and $\phi_{\rm max}=0.1 \pi$ are shown in Fig.\ref{sdprofile2}. The whole spatio-Doppler plane is discretized into $ N_d \times N_s=2N \times 5M $ grid points. Other parameters are same as those of the first example. From the images, we note that the reconstructed spatio-Doppler profiles of all considered algorithms show better performance than those when transmitting a small number of pulses (i.e., $10$) in Fig.\ref{sdprofile}. Again, we still note that the proposed algorithm exhibits similar performance with the ADMT, and much better quality than the ADM, BP-IPM and IAA algorithms. This illustrates that the proposed algorithm outperforms other algorithms, which are without array GP errors¡¯ estimation, when the target is at a low speed. Additionally, one should note that the fine characteristics of clutter spectrum are important for the sparsity-based STAP algorithms. Specifically, it might have different influences on different algorithms.

\begin{figure}[t]
\centering
  \includegraphics[width=78mm]{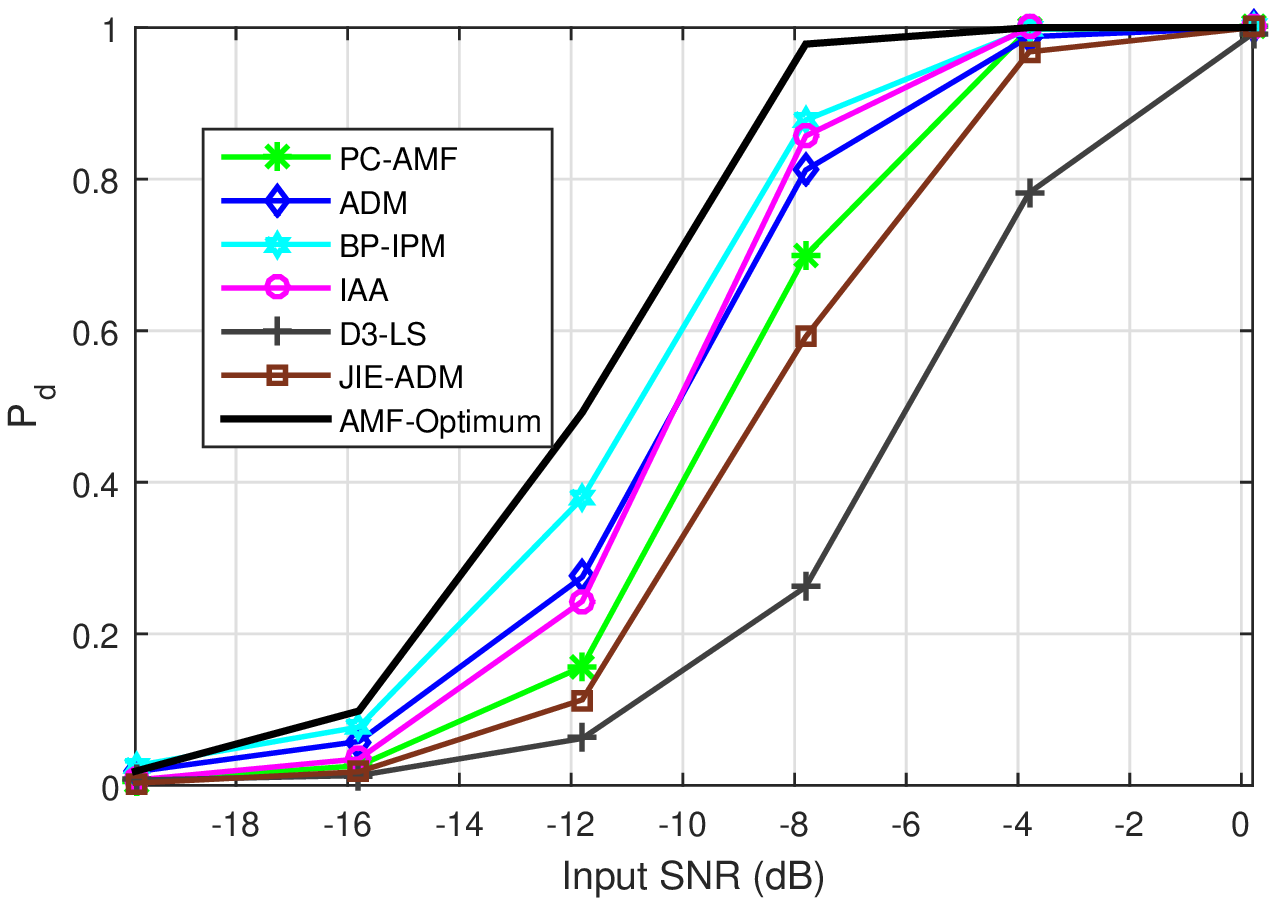}
  \caption{Impacts of the array GP errors on the detection performance of the proposed approach
and the existing ones.}\label{impacts}
\end{figure}

\begin{figure}[t]
\centering
  \includegraphics[width=78mm]{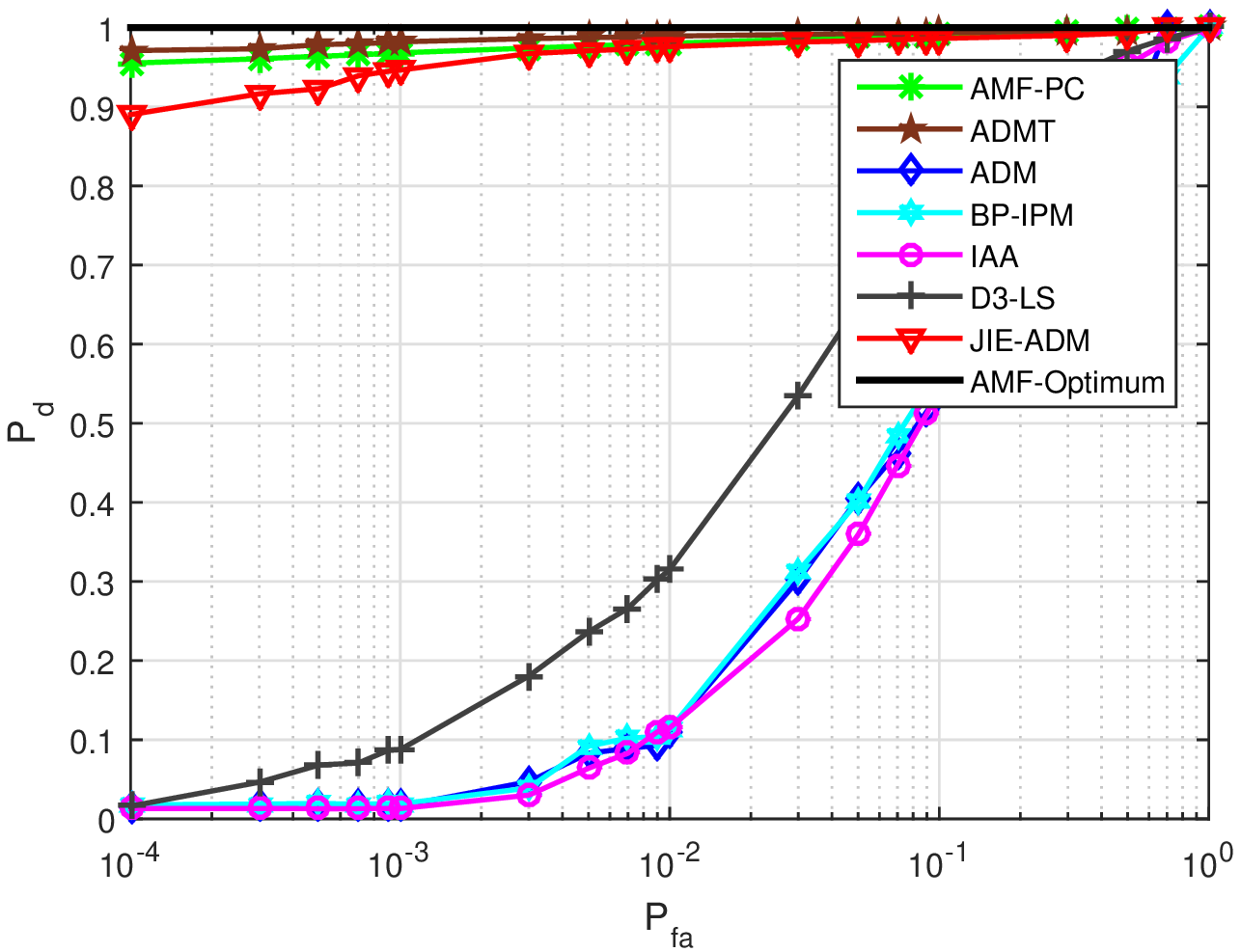}
  \caption{The ROC curves of the proposed JIE-ADM algorithm against different
Doppler frequencies at a level of array GP errors $\epsilon_{\rm max}=0.1$ and $\phi_{\rm max}=0.1 \pi$.}\label{roc}
\end{figure}

In the second example, we assess the detection performance of the proposed algorithm.
The false alarm rate $P_{fa}$ is set to $10^{-3}$, and
the target is in the boresight with the
normalized Doppler frequency $0.36$. First, in Fig.\ref{impacts},
we show the impacts of array GP errors on the detection performance of the conventional D3-LS STAP
and the existing SR algorithms, i.e., ADM, BP-IPM and IAA.
In the figure, AMF-Optimum represents the detector
of the adaptive matched filter (AMF) with clairvoyant knowledge of
the space-time covariance matrix of the interference as well
as the space-time steering vector of the target (including any ``errors").
The results in Fig.\ref{impacts} illustrate that:
(1) the detection performance of the existing SR algorithms are better than that of
the conventional D3-LS STAP when there are no array GP error,
which are coincident with conclusions in \cite{KeSun2011a,ZYangCSSP2016};
and (2) the detection performance of
the conventional D3-LS STAP and the existing SR algorithms
are significantly degraded in the presence of the array GP errors, and the proposed method achieves the best detection performance. Additionally, the performance of a typical statistical STAP
method, i.e., PC, with AMF, (namely, AMF-PC) is also shown in Fig.\ref{impacts}. The rank and the number of training snapshots used for the AMF-PC are set to $28$ and $60$, respectively. It illustrates that
the AMF-PC is not sensitive to the array GP errors. However, statistical STAP
methods require significantly more training snapshots than the sparsity-based STAP. Furthermore,
in the following simulations, when
the array GP errors increase, the performance of AMF-PC becomes worse
than the proposed algorithm (see Fig.\ref{roc}).

\begin{figure}[t]
\centering
  \includegraphics[width=78mm]{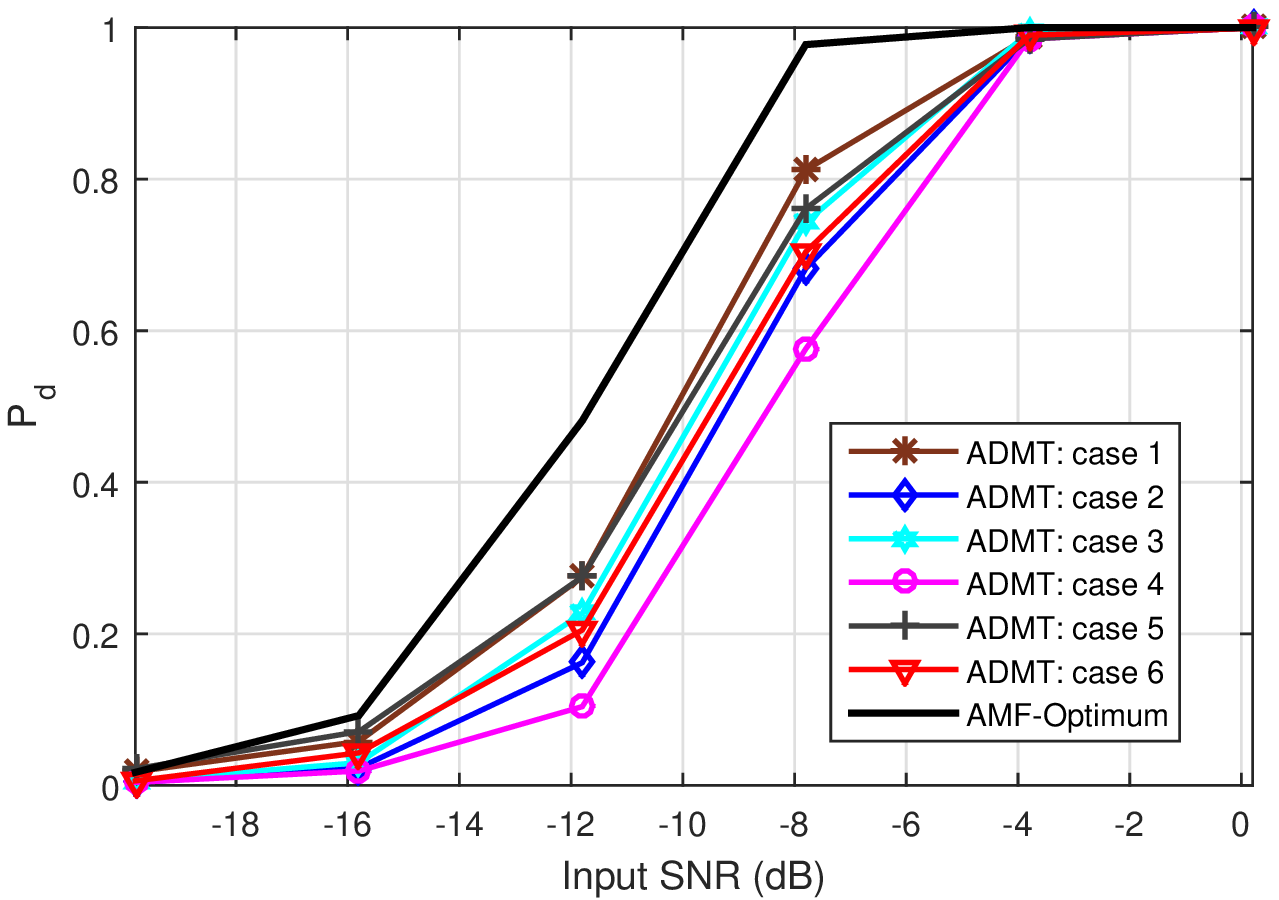}
  \caption{The detection performance of the proposed JIE-ADM algorithm and ADMT
against the target's input SNR with different cases of array GP errors: case 1, no array GP error; case 2, $\epsilon_{\rm max}=0.025$ and $\phi_{\rm max}=0.025 \pi$;
case 3, $\epsilon_{\rm max}=0.05$ and $\phi_{\rm max}=0.05 \pi$; case 4, $\epsilon_{\rm max}=0.1$ and $\phi_{\rm max}=0.1 \pi$;
case 5, $\epsilon_{\rm max}=0.15$ and $\phi_{\rm max}=0.15 \pi$; and case 6, $\epsilon_{\rm max}=0.2$ and $\phi_{\rm max}=0.2 \pi$.}\label{pd_sinr}
\end{figure}

Next, we show the detection performance of the proposed JIE-ADM
algorithm and ADMT
against the input SNR in Fig.\ref{pd_sinr}. Here, we consider six different cases of
GP errors: case 1, no array GP error; case 2, $\epsilon_{\rm max}=0.025$ and $\phi_{\rm max}=0.025 \pi$;
case 3, $\epsilon_{\rm max}=0.05$ and $\phi_{\rm max}=0.05 \pi$; case 4, $\epsilon_{\rm max}=0.1$ and $\phi_{\rm max}=0.1 \pi$;
case 5, $\epsilon_{\rm max}=0.15$ and $\phi_{\rm max}=0.15 \pi$; and case 6, $\epsilon_{\rm max}=0.2$ and $\phi_{\rm max}=0.2 \pi$.
It is seen from Figs.\ref{impacts} and \ref{pd_sinr} that
the proposed JIE-ADM algorithm provides slightly worse performance than the
ADMT, but is more robust to the array GP errors and obtains much better performance than
the conventional D3-LS STAP and existing SR algorithms. This is because
the proposed JIE-ADM algorithm provides more accurate estimate of the spatio-Doppler profile and is much more robust to the array GP errors.

\begin{figure}[t]
\centering
  \includegraphics[width=78mm]{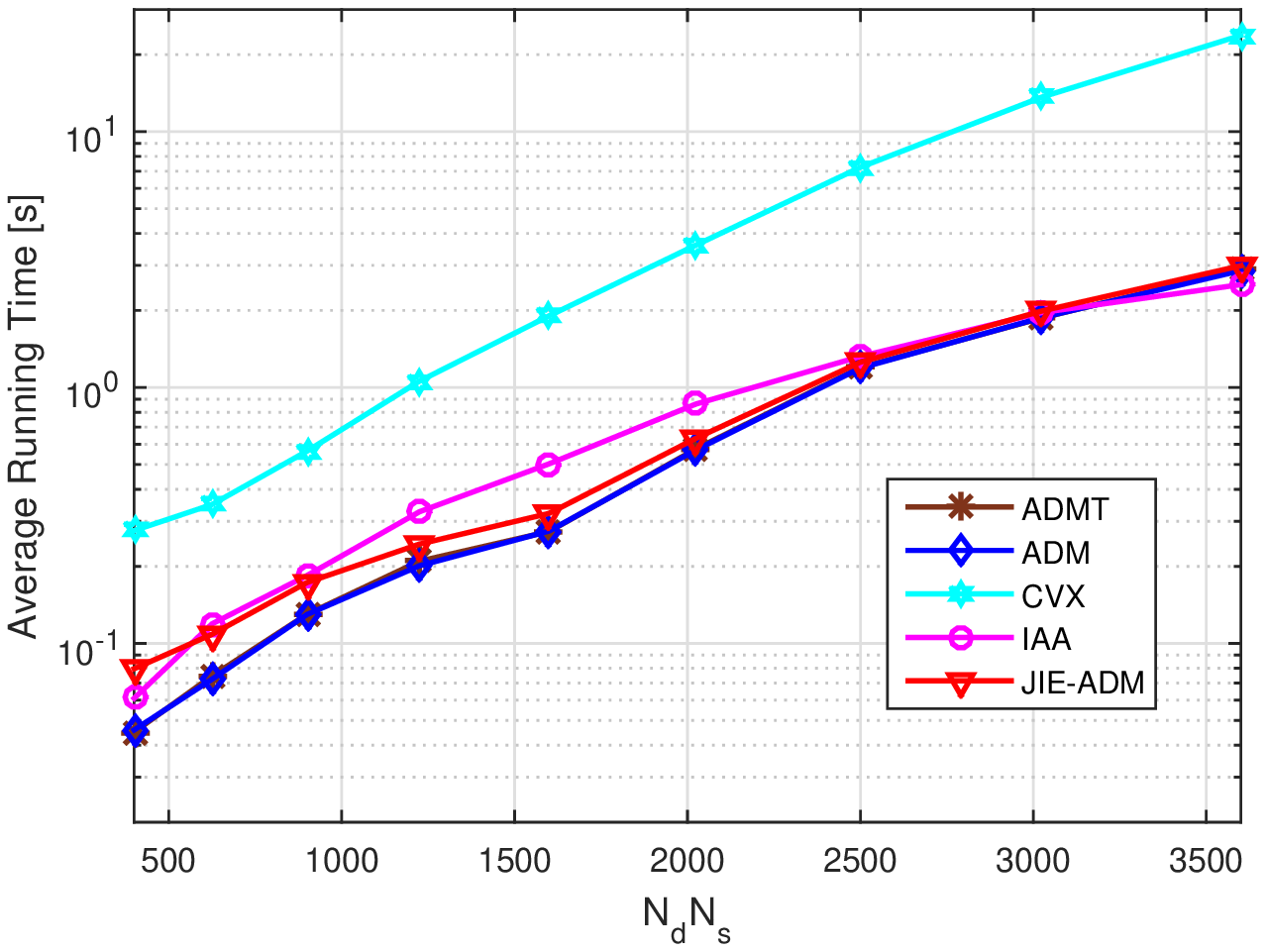}
  \caption{The average running time of the sparsity-based
STAP algorithms versus sizes of dictionary for estimating
one spatio-Doppler profile.}\label{complexity}
\end{figure}

To further investigate the performance of the proposed JIE-ADM algorithm, in Fig.\ref{roc}, we
examine the detection performance with different
Doppler frequencies at a level of array GP errors
$\epsilon_{\rm max}=0.1$ and $\phi_{\rm max}=0.1 \pi$,
by showing the receiver operating characteristic (ROC) curvers.
Here, slow, median speed and relative fast moving targets are simulated with
the normalized Doppler frequencies of $0.13$, $0.23$ and $0.36$, respectively.
The corresponding input target's SNRs are fixed to $0.2$dB, $-3.8$dB and $-3.8$dB, respectively.
The results in Fig.\ref{roc} highlight that the
proposed JIE-ADM algorithm considerably outperforms the
conventional D3-LS STAP and existing SR algorithms in presence of array GP errors
regardless of the detection of the slow, median speed or relative fast moving targets.
It should be pointed out that the detection performance of
the proposed algorithm degrades for the slow moving target.
This can be roughly understood from Fig.\ref{sdprofile} that
the difficulty to separate the target and the clutter increases
when the target is close to the clutter ridge. As
the target's input SNR increases, the detection performance improves.

Fig.\ref{complexity} plots the average running time of the sparsity-based
STAP algorithms versus sizes of dictionary for estimating
one spatio-Doppler profile. Here, the simulations are operated on a standard
desktop computer with a $3.6$GHz CPU (dual core with Matlab's multithreading
option enabled) and $4$GB of memory. The size of one CPI is changed from $16$
to $144$, corresponding to the number of columns of the dictionary from
$400$ to $3600$. The curves indicate that the computational complexity of
the proposed JIE-ADM algorithm is close to that of the ADM and ADMT algorithms.
That is to say, the added array GP errors estimation step of the proposed algorithm
costs very little, which can be also concluded from the estimation equations,
i.e., (\ref{app3}), (\ref{app4}), (\ref{app5}) and (\ref{app6}).

\section{Conclusions}
In this paper, a novel sparsity-based STAP algorithm
has been presented for airborne radar.
In order to avoid the performance degradation caused by
array errors, the proposed algorithm reformulated
the sparsity-based STAP as a joint optimization of the
spatio-Doppler profile and array errors by employing
the framework of ADM. By solving the above problem iteratively,
we developed a median CFAR detector using the
reconstructed spatio-Doppler profiles. The performance
of the proposed algorithm was tested and compared with that
of the conventional D3-LS STAP and other existing sparsity-based
STAP algorithms. Results show that the proposed algorithm
is robust to array errors and yields significant improvement
in detection performance over the conventional D3-LS STAP
and other existing sparsity-based STAP algorithms.
Additionally, the proposed algorithm adds very little
computational complexity compared with the ADM without
array error estimation. In our future work, we will investigate
fast sparsity-based STAP algorithms with jointly estimating
the spatio-Doppler profile and array errors. Moreover, the detector
design based on the spatio-Doppler profiles and its statistics will be
considered and analyzed.

\appendices
\section{Proof of (\ref{admm11})}
\label{secapp1}
Taking the gradients of the cost function in problem (\ref{admm10}) with respect to
${\bf t}^\ast$ and $\gamma^\ast$ and equating them to zeros, we have
\begin{eqnarray}\label{app1}
     {\bf Q}^H{\bf Q}{\bf t} = {\bf Q}^H{\bf z}^p + \gamma{\bf 1}_M,
\end{eqnarray}
and
\begin{eqnarray}\label{app2}
     \sum^M_{m=1}t_m = \varsigma.
\end{eqnarray}
Note that ${\bf Q}^H{\bf Q}$ is an $M \times M$ diagonal matrix, and its $m$th diagonal element is
$\sum^N_{n=1} \left|x_{(n-1)M+m}\right|^2$. Thus, substituting this into (\ref{app1}),
we obtain
\begin{eqnarray}\label{app3}
     {\bf t}^{p+1} = \left[\frac{b_1 + \gamma}{a_1}, \frac{b_2 + \gamma}{a_2}, \cdots, \frac{b_M + \gamma}{a_M}\right]^T,
\end{eqnarray}
where
\begin{eqnarray}\label{app4}
     b_m = \sum^N_{n=1} x^\ast_{(n-1)M+m}z^p_{(n-1)M+m},
\end{eqnarray}
and
\begin{eqnarray}\label{app5}
     a_m = \sum^N_{n=1} \left|x_{(n-1)M+1}\right|^2.
\end{eqnarray}
Substituting (\ref{app3}) into (\ref{app2}), we obtain
\begin{eqnarray}\label{app6}
     \gamma = \frac{\varsigma - \sum^M_{m=1}\frac{b_m}{a_m}}{\sum^M_{m=1}\frac{1}{a_m}},
\end{eqnarray}
Therefore, we have the formulation of ${\bf t}$ given in (\ref{admm11}).

\end{document}